\documentclass[oldversion]{aa}
\usepackage{txfonts}
\usepackage{graphicx}
\usepackage{amssymb}
\usepackage{epsfig}
\usepackage{natbib}
\usepackage{tabularx}
\usepackage{7358pret}
\usepackage{textcomp}
\usepackage{color}

\newcommand{\pref}{\prettyref}

\newcommand{\ebv}{$E_{B-V}$}
\newcommand{\dib}{$\lambda$}

\newcommand{\cold}[2]{$#1\times10^{#2}$ cm$^{-2}$}
\newcommand{\colds}[3]{$N({\rm #1})=#2\times10^{#3}$ cm$^{-2}$}

\newcommand{\vlsr}[1]{$v_{\rm LSR}=#1$~\kms}

\def \cma {CH$_2$CN$^-$}
\def \cm {CH$_2$CN}

\newcommand{\dbsgs}{$^1{\rm B}_1-\tilde{\rm X}~^1{\rm A}'$}

\newcommand{\be}{\begin{equation}}
\newcommand{\ee}{\end{equation}}
\newcommand{\ie}{\emph{i.e.}}
\newcommand{\eg}{\emph{e.g.}}
\newcommand{\kms}{\mbox{km\ \ensuremath{\rm{s}^{-1}}}}

\begin{document}

\title{The \cma\,molecule: Carrier of the \dib8037 diffuse interstellar band?}

\author{Martin A. Cordiner\inst{1,2}
  \and Peter J. Sarre\inst{1}
  }
\authorrunning{M.A. Cordiner and P.J. Sarre}
\titlerunning{\cma: Carrier of the \dib8037 diffuse interstellar band?}

\offprints{Martin Cordiner, \email{m.cordiner@qub.ac.uk}}

\institute{School of Chemistry, The University of
Nottingham, University Park, Nottingham, NG7 2RD, U.K. \and
Astrophysics Research Centre, School of Mathematics and Physics,
Queen's University, Belfast, BT7 1NN, U.K.}

\date{Received 23 February 2007 / Accepted 4 July 2007}

\abstract{The hypothesis that the cyanomethyl anion \cma\ is responsible for the relatively narrow diffuse interstellar band (DIB) at $8037.8\pm0.15$~\AA\ is examined with reference to new observational data.  The $0_0^0$ absorption band arising from the \dbsgs\ transition from the electronic ground state to the first dipole-bound state of the anion is calculated for a rotational temperature of 2.7~K using literature spectroscopic parameters and results in a rotational contour with a peak wavelength of 8037.78~\AA. By comparison with diffuse band and atomic line absorption spectra of eight heavily-reddened Galactic sightlines, \cma\ is found to be a plausible carrier of the \dib8037 diffuse interstellar band provided the rotational contour is Doppler-broadened with a $b$ parameter between 16 and 33~\kms\ that depends on the specific sightline. Convolution of the calculated \cma\ transitions with the optical depth profile of interstellar \ion{Ti}{ii} results in a good match with the profile of the narrow \dib8037 DIB observed towards HD 183143, HD 168112 and Cyg OB2 8a. The rotational level populations may be influenced by nuclear spin statistics, resulting in the appearance of additional transitions from $K_a$ = 1 of ortho \cma\ near $8025$ and $8050$~\AA\ that are not seen in currently available interstellar spectra.  For \cma\ to be the carrier of the \dib8037 diffuse interstellar band, either a) there must be mechanisms that convert \cma\ from the ortho to the para form, or b) the chemistry that forms \cma\ must result in a population of $K_a''$ levels approaching a Boltzmann distribution near 3~K.}


\maketitle

\section{Introduction}

The origin of the unidentified diffuse interstellar bands (DIBs) remains one of the greatest challenges in astronomical spectroscopy.  The subject has been reviewed by \citet{herbig95} and \citet{sarre06} who have highlighted research that points towards organic molecules as likely candidates for at least some of the more than $300$ DIBs.  \citet{sarre00} has put forward a hypothesis that `some, possibly many' of the diffuse interstellar bands arise from electronic transitions between ground and dipole-bound states of negatively charged polar molecules or small polar grains. It was noted that the $^rQ_0(1)$ line of the origin band of the \dbsgs\ transition of \cma\ occurs at $8037.8$~\AA, in good correspondence with the peak absorption wavelength of a diffuse interstellar band at $8037.9\pm0.3$~\AA\ \citep{galaz00}.

\section{Molecular anions as diffuse band carriers}

\subsection{Previous studies}

The electronic absorption spectrum of C$_7^-$ in the visible region was studied by \citet{tulej98} and five optical absorption bands were reported to match with the wavelengths of known DIBs. Models of diffuse cloud chemistry by \citet{ruffle99} that incorporated the desorption of seed molecules from grain surfaces were able to reproduce large abundances of C$_7^-$ (approximately $10^{-9}n_{\rm H}$), and also of C$_7$H$^-$, albeit only under a limited set of physical and chemical conditions and only for short periods of time.  The high electron affinity and density of vibrational states of C$_7$ permits rapid radiative electron attachment such that anion formation has been calculated to occur every time an electron collides with C$_7$ \citep{herbst00}.  However, the early promise of C$_7^-$ as a DIB carrier was quashed when high resolution observational and laboratory spectroscopy by \citet{galaz99}, \citet{sar00a}, \citet{lakin00} and \citet{mccall01} identified that the match between the DIBs and the wavelengths, strengths and profiles of the C$_7^-$ optical absorption bands was too poor to constitute an assignment.

Based on laboratory spectroscopy it was suggested by \citet{guthe01} that CH$_2$CC$^-$ might be a diffuse band carrier but a detailed assessment by \citet{mccall02a} showed that the wavelength match to the diffuse band at 6993~\AA\ was not acceptable and commented that the non-observation of additional $K$ sub-bands argued against an assignment to CH$_2$CC$^-$.  The \cma\ molecule has many spectroscopic characteristics in common with CH$_2$CC$^-$, both molecules being near-prolate asymmetric tops with ortho and para forms and both possessing dipole-bound excited states, but \cma\ has a closed electronic shell. 

\subsection{Anion chemistry in the ISM}

Chemistry in the diffuse ISM (with kinetic temperatures of $\sim50$~K, densities $\sim100$~cm$^{-3}$ and strong UV radiation fields) has been shown to be rich and complex by observations which include the detection of polyatomic molecules such as H$_3^+$ \citep{mccall98}, C$_3$ \citep{maier01}, and HCO$^+$, HCN, H$_2$CO, C$_2$H and $c$-C$_3$H$_2$ \citep{lucas96,lucas96b,lucas00,liszt01} at high fractional abundances similar to those observed in dark clouds.  Conventional models of gas-phase diffuse cloud chemistry struggle to explain how such high molecular abundances can be maintained in the interstellar UV radiation field.  However, as discussed by \citet{duley84} and \citet{hall95}, the availability of erosion products from carbonaceous dust grains, including polyynes (C$_n$ chains) and small molecules such as diacetylene (C$_4$H$_2$), could provide a reservoir of reagents to fuel a complex network of organic reactions in diffuse clouds and other photon-dominated regions. Large molecules should be able to resist photodissociation in the interstellar UV field by undergoing internal conversion in which the energy of absorbed photons is redistributed into the vibrational modes of the molecule due to the high density of states. 
To date the role of anions in interstellar reaction networks has generally been considered to be minimal although some consideration of their possible detection in the interstellar medium has been made \citep{sarre80}.  No reaction pathways involving anions were present in the `comprehensive' diffuse cloud models of \citet{dishoeck86} (where $\sim500$ reactions were modelled), and only a few atomic and diatomic anions are present in the UMIST 1999 database of $\sim4000$ astrochemical reactions involving $\sim400$ different species \citep{umist99}.  This is at least partly due to the lack of observational evidence for anions in space, but also reflects a common view that anions should be rapidly photoionised in the interstellar UV field.

The abundance of interstellar anions in relatively diffuse media depends in large part on the balance between radiative electron attachment \citep{dalg73} and photodetachment by the interstellar radiation field.  As discussed by \citet{herbst81} and \citet{herbst97}, a viable route for molecular anion formation in the ISM involves attachment of a free electron \emph{via} the formation of an excited temporary anion that stabilises through radiative transitions to the ground state.  Provided the radiative transition rate is high, anions should be able to form at a sufficient rate that -- relative to their neutral counterparts -- appreciable anion abundances may occur depending on the local interstellar gas pressure and radiation field.  Chemical models  of the circumstellar envelope of the carbon star IRC$+10$\textdegree$216$ by \citet{millar00} predict observable abundances of C$_n$$^-$, C$_n$H$^-$ and C$_n$N$^-$.  The discovery of the microwave signature of C$_6$H$^-$ in IRC$+10$\textdegree$216$ and TMC-1 by \citet{thaddeus06} at fractional abundances (relative to the neutral species C$_6$H) of about 2.5\% and 1\%, respectively, proves that anions arise in some regions of space in significant quantities.

\subsection{Dipole-bound states}

Anions derived from strongly polar neutral parent molecules (or grains) can support one or more `dipole-bound' electronic states at energies near to the detachment threshold and can potentially form the excited states in electronic absorption spectra in the visible region of the spectrum. Dipole-bound electronic states were first predicted by \citet{fermi47} who showed that a theoretical point dipole with a sufficiently large dipole moment ($>1.625$~D) can bind an electron in a diffuse orbital. Theoretical studies \citep{crawf70,garret78,gutsev95b,gutsev95c,gutsev95a} and experimental work \citep{moran87,brink95,lykke87,desfr96} has since demonstrated that in practice, dipole moments $\mu\gtrsim2.5$~D are required for a molecule to support at least one dipole-bound electronic state.  Laboratory studies of nitromethane \citep{compton96} and of cyanoacetylene and uracil \citep{sommer05} indicate that the presence of a near-threshold dipole-bound state can act as a `doorway' for electron capture, and that strong electronic transitions occur from the dipole-bound state to the ground state, resulting in rapid radiative stabilisation of the anion. It is possible that this effect may enhance the electron capture rate for the strongly dipolar C$_6$H molecule and therefore contribute towards the abundance of C$_6$H$^-$ observed by \citet{thaddeus06}.

\section{Spectroscopic observations and data reduction }

Medium resolution optical spectroscopic observations of early-type Galactic stars were performed using the High Resolution Echelle Spectrograph (HIRES) of the W.~M. Keck Observatory, Hawaii by G.~H. Herbig between 1995 and 1997 (private communication), who kindly made available the raw science exposures of eight heavily-reddened Galactic sightlines (shown in \pref{tab:herbig.stars}), including flat fields, Th/Ar and bias frames. Reductions were carried out using standard \textsc{iraf} echelle routines. In the region of the \dib8037 DIB special care was taken to eliminate telluric lines and fringing residuals by division with high S/N standard star spectra. The S/N of the reduced spectra is typically between 400 and 800 per pixel, with a resolving power of $42\,500$ (measured from the average of several unblended Th/Ar lines over the wavelength range of $\sim7000$ to $10\,000$~\AA). In addition, near-UV spectra from $\sim3200$ to $4000$~\AA\ were obtained for the sight-lines towards Cyg~OB2~8a, HD~168112 and HD~183143 and include spectra of a range of important interstellar species including \ion{Ca}{ii}, \ion{Ti}{ii} and various diatomic molecules that provide valuable information on the velocity distribution of the interstellar gases. The near-UV spectra typically have S/N $\sim300$ with a resolving power of $52\,500$. Around 8040~\AA\ the wavelength drift between successive arcs was 0.02~\AA, and for the near-UV observations (around 3600~\AA) the drift was 0.005~\AA.  Thus, absolute wavelength calibration errors should be $\Delta\lambda/\lambda\lesssim2.5\times10^{-6}$ (0.75~\kms) for the red/NIR and $\lesssim1.4\times10^{-6}$ (0.4~\kms) for the near-UV spectra.

\begin{table}
\centering
\begin{tabularx}{\columnwidth}{llXXXX}
\hline \hline
Target&Sp. type&$V$&\ebv&$d$ (pc)\\
\hline
Cyg OB2 5&O7 Ie&9.2&1.94&700\\
Cyg OB2 8a&O6 I&9.0&1.59&1000\\
Cyg OB2 12&B2\,I\,-\,B8\,Ia&11.4&2.8\,--\,3.4&600\\
HD 168112&O5&8.6&1.01&2000\\
HD 169034&B5 Ia&8.2&1.3&1100\\
HD 183143&B7 Ia&6.9&1.24&650\\
HD 186745&B8 Ia&7.1&0.96&1600\\
HD 229196&O5&8.6&1.22&1100\\
$\lambda$ Cyg&B5 V&4.6&0.04&50\\
$\zeta$ Peg&B8.5 V&3.4&0.00&50\\
\hline
\end{tabularx}
\caption[Heavily-reddened stars observed]{Table showing sightline data for \dib8037 observations.  For the \ebv\ values, $B-V$ photometry are from \citet{perry97} and spectral types from the modal averages of all data referenced in the SIMBAD database (URL: http://simbad.u-strasbg.fr/sim-fid.pl); intrinsic stellar photometry is from \citet{wegner94} with the exception of Cyg OB2 5 and 8a for which photometry was taken from \citet{massey91} and Cyg OB2 12 which is of uncertain spectral type and photometry as discussed by \citet{gredel94}.  Approximate stellar distances $d$ (accurate to around $\pm50$\%), were calculated from the spectral types and absolute magnitudes of \citet{wegner06} (stellar apparent magnitudes were dereddened assuming a ratio of visual to selective extinction ($R_{V}$) of 3.1).}
\label{tab:herbig.stars}
\end{table}

\section{Spectroscopy of \cma}
\label{sec:spec}

The neutral cyanomethyl radical \cm\ has a dipole moment between \emph{c.}~3.5 and 4.0~D \citep{ozeki04} and an electron affinity of \emph{c.} 1.55~eV \citep{lykke87}. The first dipole-bound state of the \emph{anion} lies $\simeq10$ meV below the ionisation continuum \citep{lykke87,moran87,gutsev95b}. The \cm\ radical is planar but on attachment of an electron the hydrogen atoms move out of plane, resulting in a $\tilde{\rm X}~^1$A$'$ electronic ground state in $C_{s}$ symmetry \citep{gutsev95b} with the mirror plane along the C-C-N backbone. In the excited $^1$B$_1$ state, the dipole-bound electron is very diffuse with only a weakly perturbative interaction with the rest of the molecule. Thus in the dipole-bound state the nuclear framework assumes a geometry which is almost identical to that of the ground-state neutral radical. The \dbsgs\ transition is of perpendicular type \citep{lykke87} with the transition dipole orientated along the $c$-axis and selection rules $\Delta{J}=\pm1$, $\Delta{K_a}=\pm1$ ($K_a$ is used although it is not strictly a good quantum number).  The rotational constants of \cma\ were determined to very high accuracy by \citet{lykke87} using high-resolution fast-ion-beam autodetachment spectroscopy. The  rotational energy level structure and some examples of \dbsgs\ transitions are shown in Figure 4. of \citet{lykke87}.

Nuclear spin statistics dictate that there exist ortho ($o$) and para ($p$) forms of \cma\ as occurs in the isoelectronic molecule cyanamide (NH$_2$CN), discussed by \citet{millen62}. If the ortho and para \cma\ abundances reflect the $o:p$ ($3:1$) nuclear spin degeneracies, transitions of ortho \cma\ would be three times stronger than equivalent transitions of the para form. It not clear whether thermalisation of the rotational energy levels of \cma\ would take place, but based on observations of other interstellar molecules, it is evident that ortho-to-para abundance ratios do not necessarily follow the nuclear spin degeneracies.  For example, according to \citet{flower84}, interconversion between ortho and para H$_2$ occurs \emph{via} collisional proton exchange in the reaction $o$-H$_2$ + H$^+\longrightarrow$ $p$-H$_2$ + H$^+$ + h$\nu$, which allows the population of the lowest $J$ levels to reach thermal equilibrium with the ISM such that $o:p$ is skewed away from the value of $3:1$.  In the case of \cma\, the photodetachment (destruction) rate has been calculated to be approximately $1.2\times10^{-8}$~s$^{-1}$ (E. Herbst \& T.~J. Millar, private communication); given the relatively slow rate of conversion between gas-phase ortho and para H$_2$ by proton exchange with H$^+$ ($3\times10^{-10}n_{\rm H^+}$ s$^{-1}$; \citet{flower84}), it follows that in the neutral ISM where H$^+$ densities are generally less than 0.1~cm$^{-3}$, collisional exchange with gas-phase \cma\ is unlikely to alter significantly the ortho-to-para population ratio.

For molecules with equivalent H atoms the rotational level populations (and $o:p$ ratios) can give clues as to the chemical reactions in which the molecule participates.  For example, if a molecule is formed by reaction with H$_2$, then the $o:p$ ratio of the product may be influenced by the $o:p$ ratio of the reagent H$_2$ (see the case of $c$-C$_3$H$_2$, studied by \citet{taka01}).  Also, according to \citet{dickens99}, the observed $o:p$ ratio of H$_2$CO, (which has the same nuclear spin statistical properties as \cma), indicates that the molecules probably formed in thermal equilibrium with cold dust grains. Both of these types of reactions could potentially influence the $o:p$ ratio of \cma.

\section{Results}
\label{sec:results}

\begin{figure}
\centering
\resizebox{0.9\columnwidth}{!}{\includegraphics{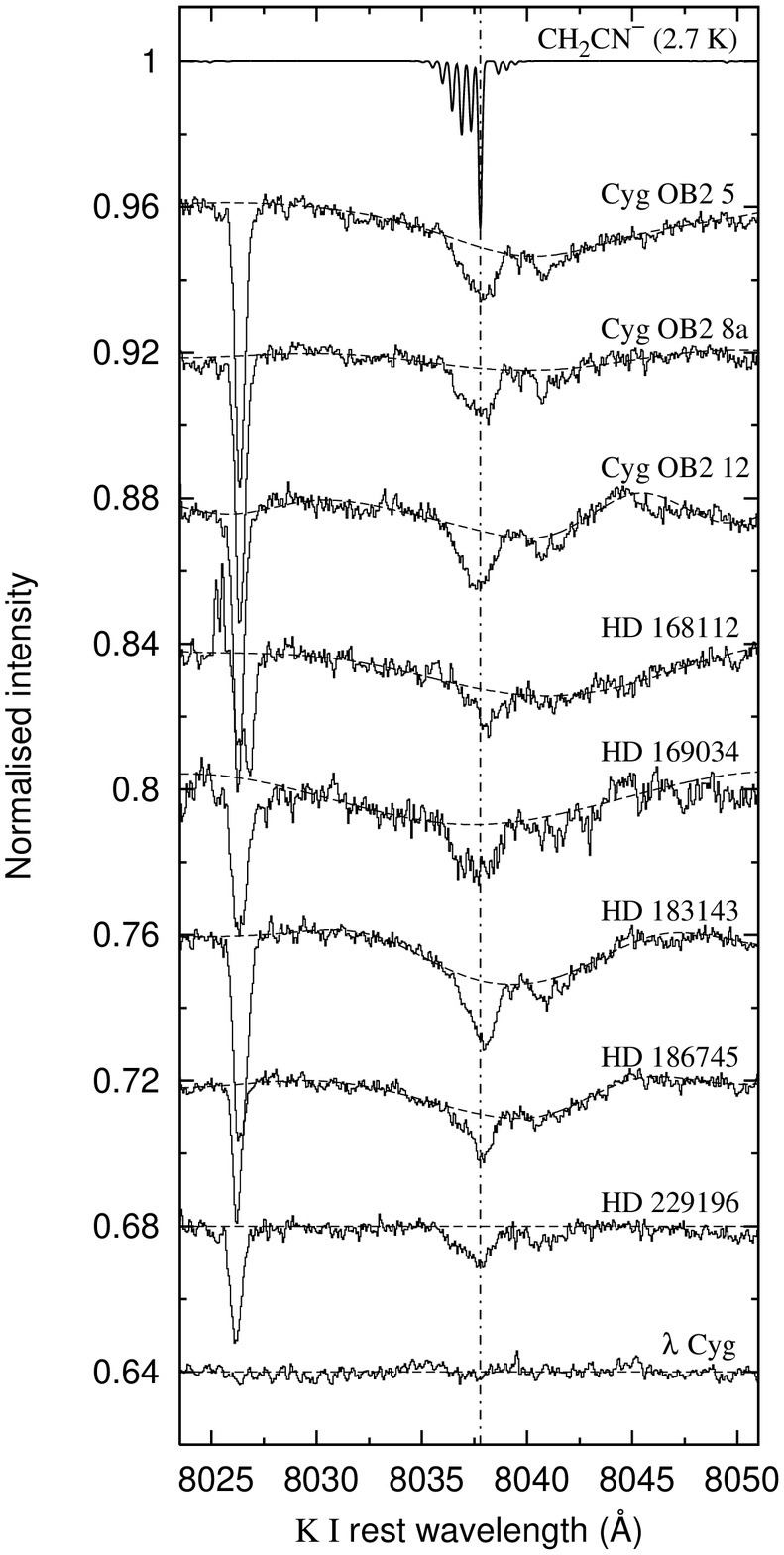}}
\caption[\dib8037 spectra compared with \cma\ \dbsgs\ calculated spectrum]{Normalised (second order Chebyshev polynomial), telluric-corrected HIRES spectra of the region around 8037~\AA. Spectra have been Doppler shifted to place the mean \ion{K}{i} wavelength at rest (using $\lambda_{\rm \ion{K}{i}\ rest}=7698.9645$~\AA\ \citep{morton03}. $\lambda$ Cyg is shown as an unreddened standard. Fitted continua are shown as dashed lines. The 2.74~K \cma\ \dbsgs\ origin band absorption spectrum is plotted at the top, calculated assuming a single cloud at rest with Doppler $b=1$~\kms\ and convolved to the resolving power $R=42\,500$ of the HIRES spectra. The $K_a''=0$ transitions lie between 8035 and 8040~\AA\ and at this temperature the $K_a''=1$ transitions at around 8025 and 8046~\AA\ are almost invisibly weak. The absorption feature at around 8026~\AA\ is an unrelated diffuse interstellar band.}
\label{fig:ch2cn.8037}
\end{figure}

Using the molecular constants of \citet{lykke87} and the \textsc{asyrot fortran} code \citep{birss84}, the \dbsgs\ transitions of CH$_2$CN$^-$ were computed for a distribution of rotational level populations in equilibrium with the 2.74~K CMB. Isoelectronic with NH$_2$CN (dipole moment 4.32~D; \citet{tyler72}), the cyanomethyl anion is expected to have a large dipole moment ($\sim4$~D) such that in the diffuse ISM the radiative transition rate is sufficient that collisional excitation of the molecule should be small.  Small changes in the level of ($J$) rotational excitation of the molecule in fact have relatively little impact on the spectrum, which is dominated by the $Q$ branch of the transition at 8037.8~\AA.  Convolved with a Gaussian with Doppler $b=1$~\kms\ and at a resolving power of $42\,500$, the calculated spectrum is displayed in \pref{fig:ch2cn.8037} for comparison with the observed HIRES spectra.

\begin{figure}
\centering
\epsfig{file=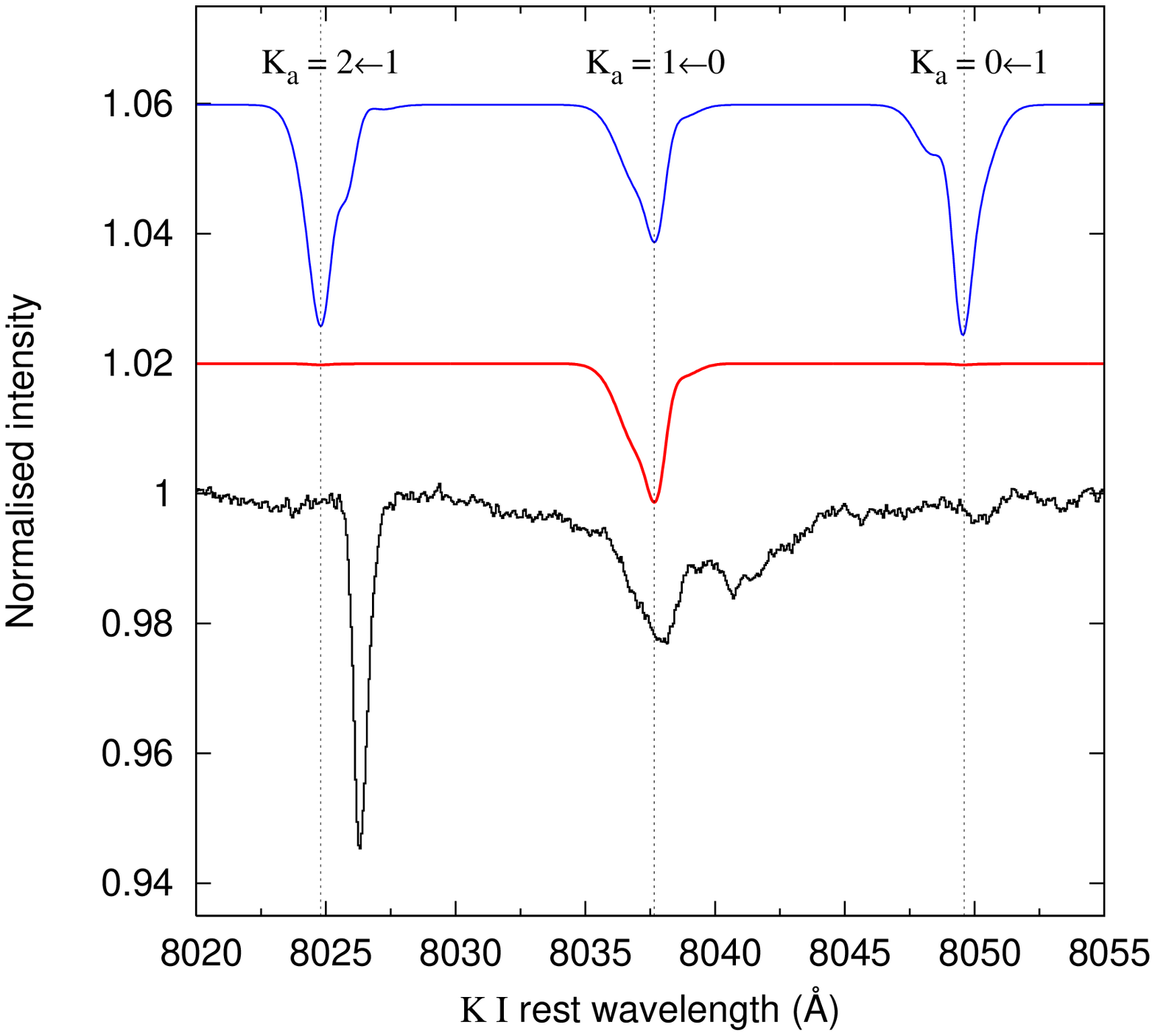, width=\columnwidth}
\caption[\dib8037 spectrum compared with \cma\ \dbsgs\ computed spectra including transitions from $K_a''=~0 and 1  $]{Comparison of computed and observed (averaged) spectra in the 8037 A region. The bottom trace shows the \emph{co-added}, normalised, telluric-corrected HIRES spectrum around \dib8037 observed towards the stars listed in \pref{tab:herbig.stars}. Spectra were Doppler-shifted to place the weighted-mean \ion{K}{i} wavelength at rest before co-addition. The \dbsgs, \{$K_a=2\longleftarrow1$, $K_a=1\longleftarrow0$, $K_a=0\longleftarrow1$\} modelled absorption spectrum is shown, and includes the transitions arising from both $K_a''=~0$ (para) and $K_a''=~1$ (ortho) \cma.  The middle spectrum was calculated assuming a thermal Boltzmann distribution of $K_a$ level populations. The upper spectrum shows the result taking the $K_a''$ populations to be determined solely by a $3:1$ nuclear spin-statistical weight ortho~:~para ratio for odd~:~even $K_a''$ levels.  The interstellar \cma\ distribution is modelled as a single cloud at rest ($v=0$~\kms) with Gaussian line-shape (Doppler $b=18$~\kms), convolved to the resolving power $R=42\,500$ of the HIRES spectra.  The respective peak absorption wavelengths of the three Doppler-broadened $K_a$ sub-bands are 8024.8, 8037.7, and 8049.6~\AA, shown as dotted lines and correspond to ortho, para and ortho \cma\ respectively.}
\label{fig:ch2cn.para}
\end{figure}

Examination of the spectrum of the standard star $\lambda$ Cyg shows no evidence for significant stellar lines, fringing artifacts or telluric residuals across the wavelength region. The most prominent features are the \dib8026 DIB, which is among the narrowest known diffuse bands, and an absorption feature around 8037~\AA. The \dib8037 DIB is relatively narrow (${\rm FWHM}\sim1.3-2$~\AA), but is overlapped by a broader absorption centered at around 8040~\AA\ \citep[noted by][]{herbig91} that is most prominent towards Cyg OB2 12 and HD 183143.  The \dib8040 component has a ${\rm FWHM}$ of around 4~\AA\ and forms an absorption peak at $8040.7\pm0.3$~\AA.  In the co-added (\ion{K}{i} rest frame) spectra (shown in \pref{fig:ch2cn.para}), Gaussian fits to the peak of the \dib8037 DIB give a wavelength of $8037.8\pm0.15$~\AA\ for this diffuse interstellar band.

The peak wavelength match of \dib8037 with the 2.7~K \dbsgs\ \cma\ spectrum shown in \pref{fig:ch2cn.8037} is very good. However, the computed spectrum clearly contains significant fine structure that is not present in any of the observed \dib8037 profiles.  The $Q$ branch of the $K_a''=0$ transition creates the prominent peak at 8037.78~\AA. The $P$ branch is of rather low intensity compared to the $R$ branch that produces the set of lines between 8035 and 8037.2~\AA.  Evidence for asymmetry of \dib8037 can be seen in the blue degradation of the DIB spectra for the three Cyg OB2 sightlines and HD~183143, HD~186745 and HD~229196.  The band shows significant evidence for profile variability across all of the sightlines, especially in comparing HD~186745 and HD~183143 with Cyg~OB2.

The variable strength and profile of \dib8040 with respect to \dib8037 suggests that it is probably caused by a different carrier. Towards Cyg~OB2~5 and HD~168112, the \dib8040 DIB has ${\rm FWHM}\sim11.5$~\AA\ -- much broader than in the other sightlines -- and perhaps indicates the presence of \emph{another} broad DIB.  To isolate the \dib8037 DIB for analysis is non-trivial among these other contaminating features. However, it is relatively narrow which assists its rejection in continuum fitting algorithms such that consistently repeatable continuum fits were possible.

\subsection{Ortho and para \cma}
\label{sec:o-to-p}

As detailed in \pref{sec:spec}, \cma\ is expected to exist in the ISM in para and ortho forms, with the lowest occupied rotational levels being $J''=0,\ K_a''=0$ and $J''=1,\ K_a''=1$, respectively.  The $K_a=1\longleftarrow0$ transitions peak around 8037.8~\AA, whereas the $K_a=0\longleftarrow1$ and the $K_a=2\longleftarrow1$ transitions peak at $8049.6$ and $8024.8$~\AA\ respectively. The \dib8037 co-added spectrum for all eight sightlines is plotted in \pref{fig:ch2cn.para} with para + ortho \cma\ \dbsgs\ transitions, convolved with a rest cloud model (\vlsr{0}) with Doppler $b=18$~\kms\ and an arbitrary intensity scaling.  Two $K_a''$ population scenarios are plotted: (1) a $3:1$ ratio of odd~:~even $K_a''$ level populations and (2) a 2.74~K Boltzmann distribution.

The predicted $K_a=2\longleftarrow1$ transitions (with peak absorption at $\sim8024.8$~\AA) partially overlap the narrow \dib8026 DIB.  There is some evidence for a small blue shoulder on this DIB, at a wavelength slightly blue of the peak \cma\ $K_a=2\longleftarrow1$ absorption wavelength.  The $K_a=0\longleftarrow1$ feature (with peak absorption at $\sim8049.6$~\AA) falls close to a small absorption feature in the co-added spectrum.  In this spectrum, the central depth of the \dib8037 feature is $1.35\pm0.16$\% and the central depth of the \dib8049 feature is $0.3\pm0.16$\% ($2\sigma$ error estimates derived from the RMS noise of the continuum).  The ratio of the central depth of the \dib8037 to \dib8049 features is consistent with a $o:p$ ratio of $\sim1:2.3$, but given the mismatch between the calculated and observed profile and peak wavelength, it seems unlikely that \dib8049 is caused by $o$-\cma.

From \pref{fig:ch2cn.para}, it is clear that if interstellar \cma\ has a $3:1$ `statistical' ratio of $o:p$ states, it cannot be the carrier of the \dib8037 DIB due to the lack of the ortho transitions at the expected strengths in the observed spectra.  The \dib8026 DIB is too red relative to the $Q$ branch (main peak) of the $K_a=2\longleftarrow1$ transition to constitute a spectroscopic match with the model calculation.  If however the $K_a$ level populations have, or approach, a Boltzmann distribution, then the transitions originating in $K_a=1$ are sufficiently weak for \cma\ to be a plausible carrier of the \dib8037 diffuse interstellar band.

\begin{figure}
\centering
\epsfig{file=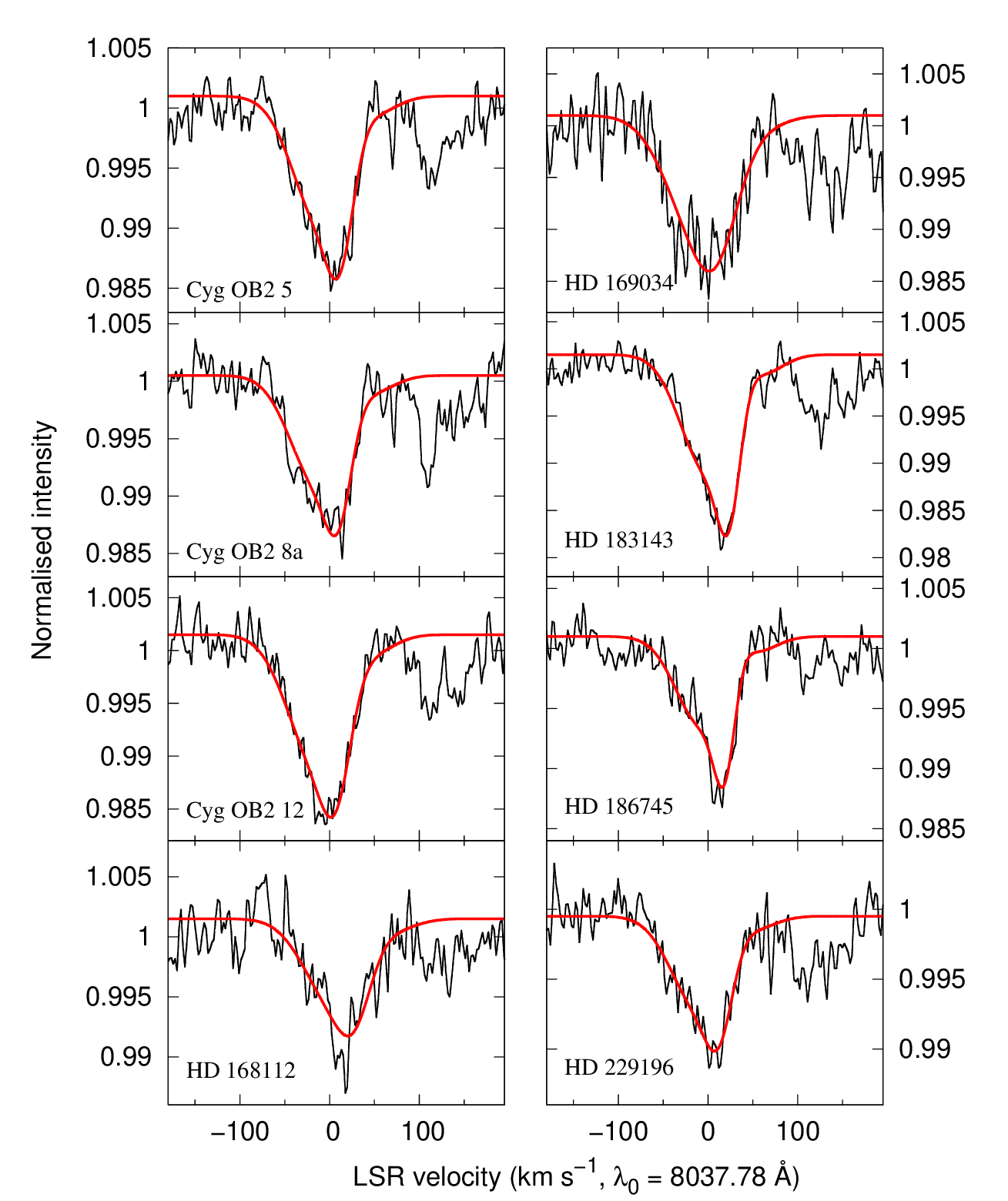, width=\columnwidth}
\caption[Interstellar \cma\ \dib8037 fits]{Telluric-corrected, normalised DIB \dib8037 spectra (thin black traces) and least-squares fitted \cma\ \dbsgs, $K_a=1\longleftarrow0$ models (thick red traces).  Models employ a single Gaussian interstellar cloud. Radial velocity shift, Doppler width, central depth and an additive continuum offset were free parameters in the fits.  The DIB rest wavelength was set at $\lambda_{0}=8037.78$~\AA\ for the displayed velocity scale. Best-fitting model parameters are shown in \pref{tab:ch2cn.dib.pars} and the inferred \cma\ interstellar velocity distributions plotted in \pref{fig:species} (lines labeled `Mdl').}
\label{fig:ch2cn.fits}
\end{figure}

\begin{table}
\centering
\begin{tabularx}{\columnwidth}{lXXXXX}
\hline\hline
Sightline&$v_{\rm LSR}$&$b$&$W_{8037}$&$\sigma_{c}$&$\sigma_{f}$\\
\hline
Cyg OB2 5&13.9&21.6&32.5&0.0020&0.0014\\
Cyg OB2 8a&12.3&21.0&30.2&0.0024&0.0016\\
Cyg OB2 12&9.4&24.0&38.3&0.0024&0.0013\\
HD 168112&29.9&26.4&24.5&0.0018&0.0021\\
HD 169034&13.6&33.0&38.7&0.0036&0.0024\\
HD 183143&25.0&18.0&37.5&0.0022&0.0012\\
HD 186745&19.3&16.2&24.3&0.0014&0.0013\\
HD 229196&14.1&21.6&22.0&0.0020&0.0012\\
\hline
\end{tabularx}
\caption[Interstellar \cma\ \dib8037 fit parameters]{Least-squares fit parameters of the \dib8037 \cma\ models shown in \pref{fig:ch2cn.fits}. The mean LSR velocity ($v_{\rm LSR}$~/~\kms), Doppler cloud width ($b$~/~\kms) and equivalent width of the model \dib8037 DIB ($W_{8037}$~/~m\AA) are given along with the normalised continuum RMS of the observed spectra ($\sigma_c$), and the RMS of the model fit residuals ($\sigma_f$).}
\label{tab:ch2cn.dib.pars}
\end{table}

\begin{figure}
\centering
\epsfig{file=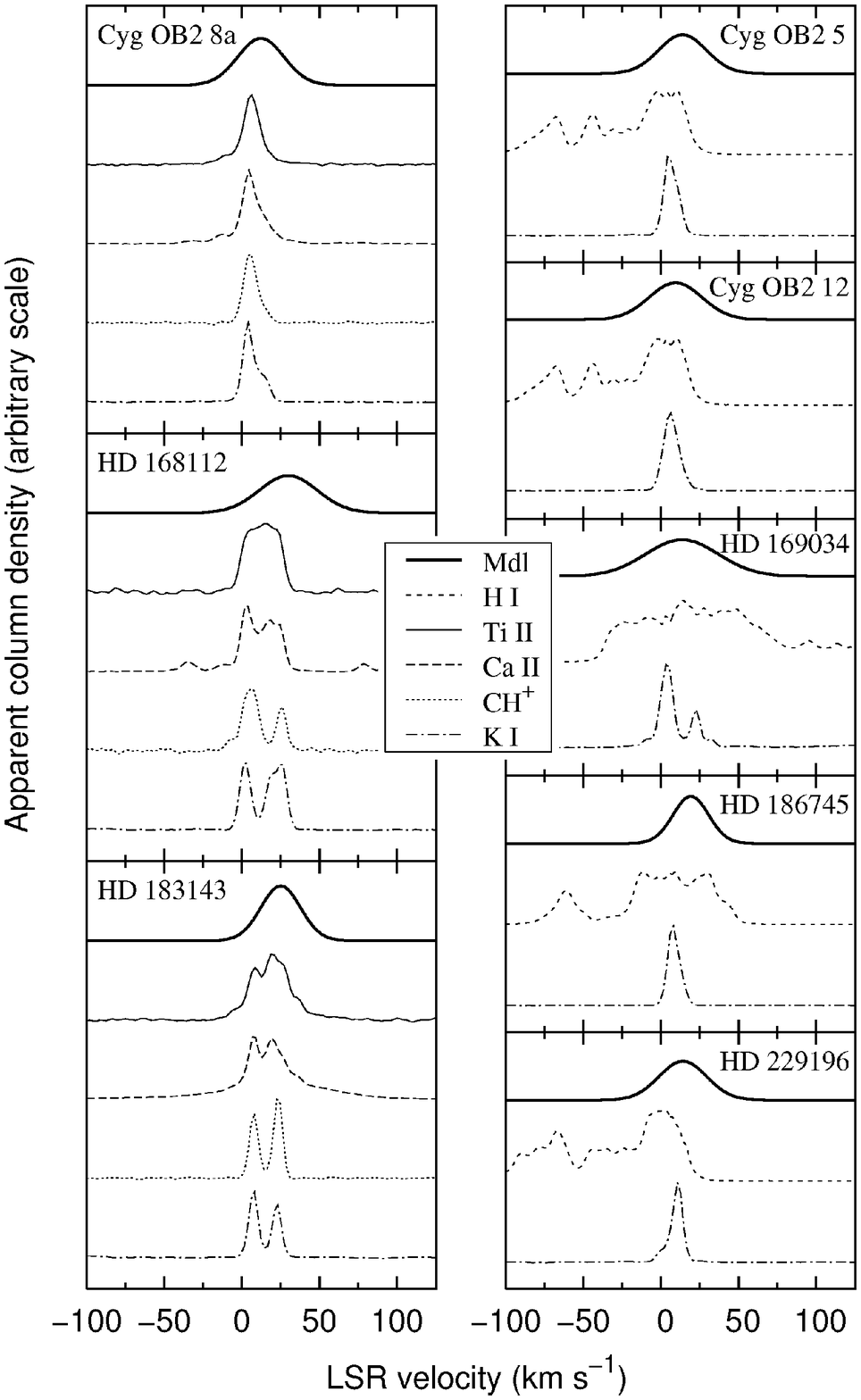, width=\columnwidth}
\caption[Interstellar species]{Each of the eight panels above shows LSR velocity-space apparent column density profiles for interstellar species for which data were available.  Data for the \ion{K}{i} 7699~\AA, CH$^{+}$ 3958~\AA, \ion{Ca}{ii} 3968~\AA\ and \ion{Ti}{ii} 3384~\AA\ lines are shown.  Hypothetical model \cma\ velocity distributions (labeled `Mdl') are shown at the top of each panel, calculated based on the fits shown in \pref{fig:ch2cn.fits} (parameters given in \pref{tab:ch2cn.dib.pars}) with the assumption that the narrow \dib8037 DIB is caused by the origin band \dbsgs\ transitions of \cma\ at 2.74~K, and that the interstellar velocity distribution of the gas is Gaussian. The interstellar \ion{K}{i}, CH$^{+}$, \ion{Ca}{ii} and \ion{Ti}{ii} apparent column densities ($N_a$) were calculated from the normalised HIRES spectra ($I/I_0$, where $I_0$ is the continuum intensity) according to $N_a\propto\ln(I_0/I)$ \citep[see for example][]{savage91}. For the right-hand panel, \ion{H}{i} column density profiles are from the Leiden/Argentine/Bonn (LAB) Survey of Galactic \ion{H}{i} \citep{lab05}.}
\label{fig:species}
\end{figure}

\begin{figure}
\centering
\epsfig{file=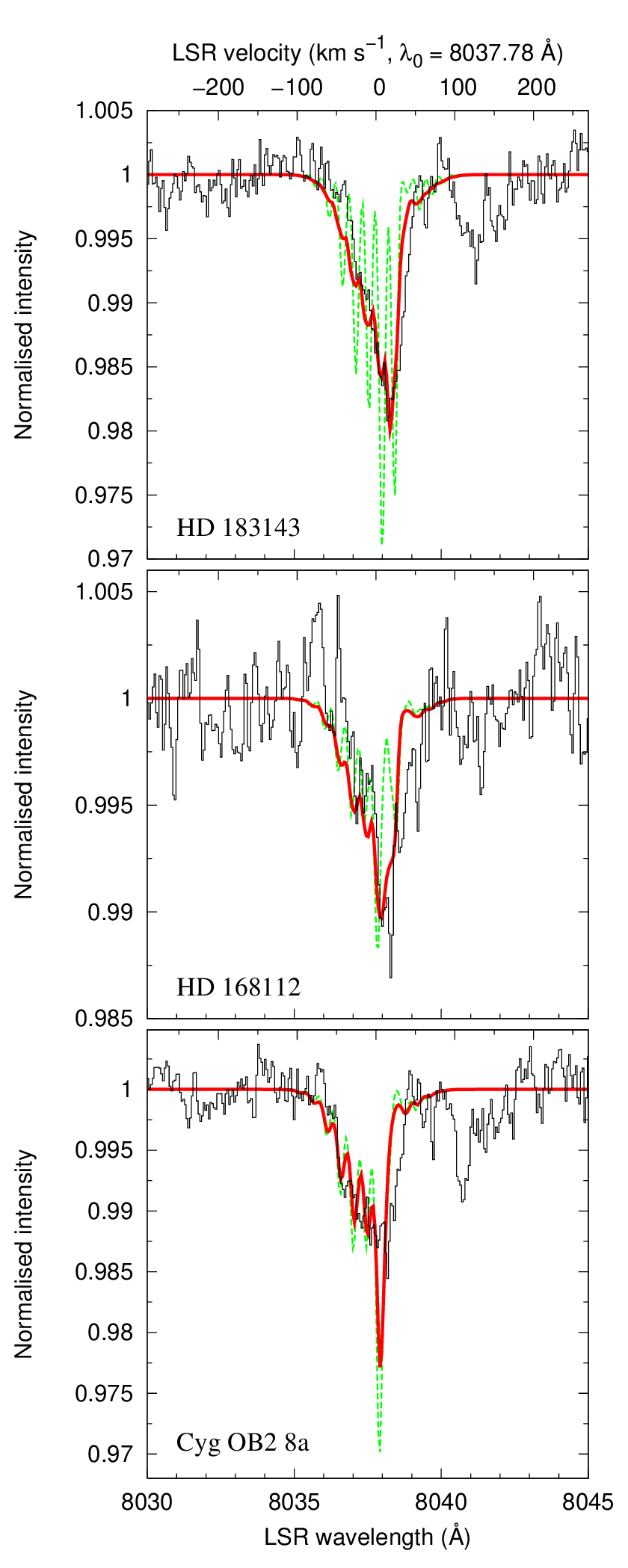, width=0.9\columnwidth, height=0.87\textheight}
\caption[]{2.74~K \cma\ \dbsgs, $K_a''=0$ transitions convolved with the HIRES PSF and the interstellar \ion{Ti}{ii} profiles (solid lines), and the \ion{K}{i} profiles (dashed lines), observed towards HD~183143, HD~168112 and Cyg~OB2~8a, overlaid on the normalised, telluric-corrected \dib8037 DIB profiles (thin black histograms).}
\label{fig:convolutions}
\end{figure}

\subsection{Modelling the \dib8037 diffuse interstellar band}

As shown by Figures \ref{fig:ch2cn.8037} and \ref{fig:ch2cn.para}, a large Doppler broadening of the interstellar gas is required for \cma\ to be the carrier of \dib8037.  Based on the assumption that the \dib8037 DIB is caused by the origin band \dbsgs, $K_a''=0$ transitions of \cma\ at 2.74~K, it is possible to use the shape of the DIB to infer the shape of the hypothetical interstellar \cma\ velocity distribution. This distribution may then be compared with observed interstellar velocity profiles for different species in the observed sightlines.  The computed 2.74~K absorption spectrum was subjected to a least-squares analysis whereby it was convolved with a single Gaussian interstellar cloud model with the Doppler width, mean radial velocity and equivalent width floated as free parameters.  A fourth free parameter allowed an additive shift in the continuum height if necessary. The least-squares parameters for each sightline are shown in \pref{tab:ch2cn.dib.pars}, and the corresponding DIB model spectrum and \cma\ velocity distribution are shown in Figures \ref{fig:ch2cn.fits} and \ref{fig:species}, respectively.  The quality of the fits was monitored by comparing the normalised continuum RMS of the observed spectra ($\sigma_c$) with the normalised RMS of the model fit residuals ($\sigma_f$).

\subsection{Hypothetical interstellar \cma\ distribution}

As shown by \pref{fig:ch2cn.fits}, the model fits match the spectra to a high degree of accuracy for each of the eight sightlines.  Variations in the width and structure of the DIB profile can be reproduced in each case, highlighted by the progression of the \dib8037 ${\rm FWHM}$ from narrow to wide in the sequence HD~186745 $\rightarrow$ HD~183143 $\rightarrow$ Cyg~OB2~12 $\rightarrow$ HD~169034.  The progression corresponds to a sequential increase in the Doppler $b$ of the cloud model: 16.2 $\rightarrow$ 18.0 $\rightarrow$ 24.0 $\rightarrow$ 33.0~\kms.  Such large $b$ values would require unphysically large kinetic and/or turbulent gas velocities and clearly cannot arise in a single interstellar cloud, and this places some doubt on the validity of the single-cloud model used here.  However, given the S/N of the spectra, there is no statistical justification for employing more than one cloud in the model. The Gaussian cloud model fits the DIB well as shown by $\sigma_c\approx\sigma_f$, implying that if \cma\ is the \dib8037 carrier, its interstellar velocity distribution must be approximately Gaussian.

Comparing the hypothetical \cma\ velocity structures with the profiles of interstellar lines in \pref{fig:species} shows that the calculated \cma\ distributions are reasonably close to the velocities of the diffuse interstellar clouds.  Due to the lack of spectroscopic data for other interstellar species for five of the sightlines examined, \ion{H}{i} data were obtained from the Leiden/Argentine/Bonn (LAB) survey \citep[][data available at http://www.astro.uni-bonn.de/$\sim$webrai/english/tools\_labsurvey.php]{lab05}.  Although background \ion{H}{i} emission contaminates these profiles, they at least indicate the likely velocity structure of a significant proportion of the neutral gas in these sightlines, aided by the fact that the stars are relatively distant (see \pref{tab:herbig.stars}). The peak of the calculated \cma\ distribution generally lies within c. 10~--~15~\kms\ of the peak \ion{H}{i} column density. It should be noted that the mean LSR velocities of the foreground clouds traced by \ion{K}{i} are positive for each sightline, which can be explained as a result of Galactic rotation and observational bias in the Galactic longitude of the targets. In \pref{fig:species} there is an additional systematic small shift in the inferred \cma\ distribution to a higher LSR velocity as compared to the \ion{K}{i}. It is possible that this originates from the difficulty in defining the continuum of $\lambda$8037 as it sits atop a (variable) 8040~\AA\ broader feature.

Analysis of the velocity structure of interstellar gas in the observed sightlines has been performed to determine whether the DIB profiles can be reproduced by convolving the transitions of \cma\ with the observed interstellar gas distributions. The results of the least-squares fitting (\pref{tab:ch2cn.dib.pars}) show that a Doppler $b$ parameter of $\sim20$~\kms\ is required to achieve a good fit with the DIB.  Typically, clouds traced by \ion{K}{i} are much narrower than this \citep{welty01}, and even in heavily-reddened lines of sight such as those observed, the presence of multiple \ion{K}{i} clouds separated in velocity space is insufficient to account for such broadening.  Examining the data in left-hand panel of \pref{fig:species}, the HD~183143 least-squares model \cma\ distribution shows gross similarities with the \ion{Ti}{ii} and \ion{Ca}{ii} profiles.  It is evident from these spectra that \ion{Ti}{ii} has a broader and less-peaked distribution than the other species. In order of decreasing profile width and increasing structure, \ion{Ti}{ii} comes first followed by \ion{Ca}{ii}, then CH$^{+}$, and finally \ion{K}{i}. This sequence is believed to reflect the typical conditions in the clouds in which these species are generally found \citep[see for example][]{jenk89,crink94,welty98,welty03,pan05}. \ion{Ti}{ii}, due to its high condensation temperature tends to be found in the gas phase in hotter, more heavily shocked higher velocity clouds, and is depleted out onto grains in cool, quiescent diffuse clouds and molecular clouds. Ca has a similar condensation temperature to Ti and therefore comes out of the solid state to produce \ion{Ca}{ii} in similar energetic/shocked regions to where \ion{Ti}{ii} is observed, but will be ionised to \ion{Ca}{iii} in more diffuse, strongly irradiated clouds, so it tends to be found in greatest abundances in warm yet moderately well-shielded regions. CH$^{+}$ is photodissociated in less dense regions, but has been shown to have a broader velocity distribution than \ion{K}{i} \citep{pan05}, consistent with the formation of this molecule in warmer haloes surrounding cold cloud cores \citep{crawf94}. \ion{K}{i} is a good tracer for cool, quiescent, well-shielded clouds with low thermal broadening where the radiation field is attenuated and the photoionisation rate is relatively low. Based on this information and the breadth of the \ion{Ti}{ii} distributions shown in \pref{fig:species}, gas traced by \ion{Ti}{ii} may plausibly have a velocity distribution sufficient to provide the Doppler broadening required for \cma\ to be the carrier of \dib8037.  \ion{Ti}{iii} and other gases present in regions where \ion{H}{ii} is found may exist over an even greater range of velocities.  

Figure \ref{fig:convolutions} shows the convolution of the 2.74~K \cma\ \dbsgs, $K_a''=0$  transitions with the (HIRES) instrumental PSF and with the interstellar \ion{Na}{i} and \ion{Ti}{ii} profiles for each of the three sightlines. Noiseless \ion{Na}{i} and \ion{Ti}{ii} velocity profiles were derived by modelling the \ion{Na}{i} UV doublet and the \ion{Ti}{ii} 3242 and 3384~\AA\ lines using the \textsc{vapid} routine \citep{howarth02}.  The accuracy of the derived Doppler $b$ parameters was improved by the simultaneous modelling of two transitions of each species (originating in the same ground state), with differing oscillator strengths. Due to its similar chemistry, \ion{Na}{i} traces the same type of interstellar material as \ion{K}{i}, and was used in preference to the \ion{K}{i} 7699~\AA\ lines shown in \pref{fig:species} due to the reduced saturation and telluric contamination of the \ion{Na}{i} UV doublet. Atomic transition parameters were taken from \citet{morton03}.

Within the spectral signal-to-noise, the profiles in \pref{fig:convolutions} calculated by convolution of the \cma\ transitions with the interstellar \ion{Ti}{ii} profiles compare favourably with the diffuse interstellar band wavelength and profile, especially for HD~168112 and HD~183143.  Convolution by \ion{Ti}{ii} gives a significantly better fit than convolution by \ion{Na}{i} in all three cases, where the narrowness of the \ion{Na}{i} Doppler spread produces structure in the resulting contour which is not seen in the observed DIB. These results demonstrate that the origin band \cma\ \dbsgs, $K_a=1\longleftarrow0$ transitions at 2.74~K are capable of reproducing the narrow \dib8037 DIB towards HD~183143 and HD~168112 provided that the postulated \cma\ carrier molecule is distributed in velocity space in approximately the same way as interstellar \ion{Ti}{ii}. Due to the lack of substructure in the observed \dib8037 profile, it is very unlikely that \cma\ could instead coexist with \ion{Na}{i} and \ion{K}{i} in the cold neutral medium. For Cyg OB2 8a the peak absorption wavelength and profile match is reasonably good for the case of both \ion{Na}{i} and \ion{Ti}{ii}, but there is clearly a significant level of structure in the calculated profiles that is not present in the observation. For the \ion{Ti}{ii}-convolved calculations, the relatively small differences between the calculated and observed profiles could plausibly be the result of Poisson noise of the spectrum or other features contaminating the DIB profile such as other interstellar features, telluric division or flat-fielding residuals.  

\section{Discussion}

Under the assumption that the \cma\ distribution is the same as that of \ion{Ti}{ii} -- with the implication that \cma\ co-exists with gas-phase \ion{Ti}{ii} -- the fit to the observed \dib8037 spectrum towards HD~183143 and HD~168112 is very good (and to a lesser degree for Cyg~OB2~8a).  This result leads to two possibilities for the type of gas with which the postulated \cma\ might be most closely associated: (1) The warm, shocked `intercloud' medium; for the $\sim50$ interstellar clouds expected to be present in the heavily-reddened sightlines studied \citep[see][]{welty03}, the warm, low density, shocked clouds with low depletions should dominate the \ion{Ti}{ii} profiles \citep[\eg][]{crink94}. (2) The neutral ISM; \ion{Ti}{ii} is a good tracer of neutral hydrogen due to its ionisation potential of $13.6$~eV \citep{stokes78}.  Without knowledge of the precise \ion{H}{i} velocity profiles, the possibility that the postulated \cma\ co-exists with neutral atomic hydrogen gas cannot be ruled out.

There are no reports of past attempts to detect \cma\ or \cm\ in the diffuse ISM. The neutral radical \cm\ has been observed in the dense TMC-1 and Sgr B2 molecular clouds \citep[see][]{irvine88,turner90} through its pure rotational transitions. Fractional abundances are $\sim10^{-9}$ to $10^{-11}$ relative to H$_2$, and \cm\ was found to be similarly distributed to CH$_3$CN and C$_4$H. The synthesis of \cm\ in the shocked diffuse interstellar clouds traced by \ion{Ti}{ii} could be driven by the release of hydrocarbons into the gas phase during supernova shock-induced collisions between carbonaceous grains \citep{duley84,hall95}. Given a sufficient abundance of ${\rm C}_2{\rm H}_4^+$ in the gas phase, the reaction

\be
{\rm C}_2{\rm H}_4^+ + {\rm N} \longrightarrow {\rm C}{\rm H}_3{\rm CN}^+ + {\rm H},
\label{eq:ch2cn.prod}
\ee

with a rate coefficient of $\sim10^{-10}$~cm$^3$~s$^{-1}$ \citep{herbst90}, could proceed rapidly enough to produce sufficient quantities of ${\rm C}{\rm H}_3{\rm CN}^+$ to allow dissociative recombination (DR)

\be
{\rm C}{\rm H}_3{\rm CN}^+ + e \longrightarrow {\rm CH}_2{\rm CN} + {\rm H}
\label{eq:ch2cn.dr}
\ee

to occur.  This mechanism for \cm\ production is only viable if single H-atom loss is a significant channel in the DR of ${\rm C}{\rm H}_3{\rm CN}^+$ (a process which is yet to be studied in detail), and if recombination occurs before ${\rm C}{\rm H}_3{\rm CN}^+$ is destroyed by reacting with neutral hydrogen.  The dominant mechanisms by which \cm\ and ${\rm CH}_3{\rm CN}$ are expected to be destroyed in the diffuse ISM are by photodissociation and by reaction with C$^+$.

The most obvious pathway for production of the cyanomethyl anion \cma\ is by radiative electron attachment to \cm:

\be
{\rm CH}_2{\rm CN} + e \longrightarrow {\rm CH}_2{\rm CN}^- + h\nu,
\label{eq:rad.att}
\ee

though an alternative route exists \emph{via} dissociative attachment to ${\rm CH}_3{\rm CN}$:

\be
{\rm CH}_3{\rm CN} + e \longrightarrow {\rm CH}_2{\rm CN}^- + {\rm H}.
\ee

The last step was studied by \citet{sailer03} in the laboratory.  The dissociative electron attachment cross-section peaked at $4\times10^{-23}$ m$^2$ for electron energies of 3.2~eV, which provides an efficient route to the formation of \cma\ from ${\rm CH}_3{\rm CN}$.  The cyanomethyl anion production channel is $\sim500$ times more efficient than those of the other possible anionic fragments (CHCN$^-$, CCN$^-$, CN$^-$ and CH$_3$$^-$) for electrons energies between 1 and 10~eV. In the interstellar context, photon and cosmic ray impact on grains could provide a source of electrons with a few eV energy.

If the dominant chemical reactions involving \cma\ are assumed to be formation of the anion \emph{via} radiative attachment (\pref{eq:rad.att}) and destruction \emph{via} photodetachment, then in equilibrium, the anion-to-neutral ratio is given by 

\be
\frac{[{\rm CH_2CN^-}]}{[{\rm CH_2CN}]}=\frac{\alpha n_e}{\Gamma}
\label{eq:anion.frac}
\ee

where $\alpha$ is the rate of radiative electron attachment, $n_e$ is the electron density and $\Gamma$ is the photodetachment rate. Following s-wave electron capture by \cm\ into an excited state of \cma\ \citep{herbst00}, if subsequent radiative relaxation occurs rapidly, either by vibrational transitions of the vibrationally excited ground state or by electronic transitions from the dipole-bound state to the ground state, then $\alpha$ can be estimated using Equation (6) of \citet{herbst00}. In the diffuse ISM (at $T=100$~K), this yields a value of $\alpha=2.2\times10^{-7}$ cm$^{-3}$\, s$^{-1}$, assuming that the exited anion undergoes radiative stabilisation before the electron detaches back to the continuum. The photodetachment rate $\Gamma$ has been calculated to be approximately $1.2\times10^{-8}$ s$^{-1}$ (E. Herbst \& T.~J. Millar, private communication) in the standard interstellar radiation field \citep{draine78}.  Both parameters are uncertain due to the lack of laboratory or theoretical data for this molecule, but nevertheless permit a rough estimate of the fraction of \cm\ likely to occur in anionic form.  Using an electron density in the diffuse ISM of between 0.04 and 0.23 cm$^{-3}$ \citep{welty03}, \pref{eq:anion.frac} yields an anion-to-neutral ratio between 0.7 and 4.2. If \cma\ is present in the warm neutral medium traced by \ion{Ti}{ii} (and \ion{Ca}{ii}) where $T\sim4000$~K \citep[see][]{welty96} then $\alpha=3.4\times10^{-8}$~cm$^{-3}$\, s$^{-1}$ and the calculated anion-to-neutral ratio is 0.1~--~0.7.

A further test of the plausibility of \cma\ as carrier of the \dib8037 DIB is whether a feasible \cm\ abundance can be deduced from the equivalent width of the observed DIB. Polyatomic molecules are likely to be present in all of the heavily-reddened sightlines towards which \dib8037 was observed.  Cyg OB 12, Cyg~OB2~5 and HD~183143 have been studied by \citet{mccall98,mccall02b}, particularly with reference to interstellar H$_3$$^+$, detected in all three sightlines with Doppler $b\sim10$~\kms\ which is substantially broader than the $^{13}$CO and HCO$^+$ lines observed towards Cyg~OB2~12. The HCO$^+$ column density was found to be $\sim10^{11}$~cm$^{-2}$.  \citet{scapp02} determined a likely total number density of $n\gtrsim10^4$ cm$^{-3}$ for the gas containing the CO and HCO$^+$, with the inference of dense, compact molecular clumps within a more diffuse medium. The approximate molecular hydrogen column density towards Cyg OB2 12 \citep[from][]{mccall02b} is \cold{6.5}{21}\ such that the fractional HCO$^+$ abundance is  $\sim1.5\times10^{-11}n_{\rm H_2}$.

The equivalent width of the \dib8037 DIB is 38.3~m\AA\ (see \pref{tab:ch2cn.dib.pars})
which requires a column density of \colds{CH_2CN^-}{6.7}{10}$/f$ where $f$ is the oscillator strength of the transition.  Transitions between the ground and dipole-bound states are strongly observed in the lab~\citep{lykke87}, suggesting that the value of $f$ should be relatively large. Therefore, if $f=0.5$ and assuming a \cm\ anion-to-neutral ratio of 1, a molecular fraction of \cm\ of $2\times10^{-11}n_{\rm H_2}$ is required towards Cyg~OB2~12 to produce the observed \dib8037 equivalent width.  Thus, provided the \dbsgs\ transition is strong, the required interstellar \cm\ fractional column density is comparable to that of HCO$^+$ observed in the Cyg OB2 12 sightline.  Averaged along the sightline, the required interstellar \cm\ fractional abundance is at the lower end of the range of values found in dense molecular clouds ($\sim10^{-11}-10^{-9}n_{\rm H_2}$).  If diffuse and dense cloud molecular abundances are similar, then enough \cma\ can probably be produced to create the \dib8037 DIB along heavily-reddened sightlines.

The spectra observed by \citet{mccall02b} show clearly that the interstellar material traced by H$_3^+$ has a broader velocity distribution than the molecules (CO, C$_2$, CN) which typically trace dense molecular material. The H$_3^+$ line FWHM observed towards Cyg~OB2 12, Cyg~OB2 5 and HD~183143 range from 8 to 15~\kms\ whereas the CN line FWHM are all less than 4~\kms. \citet{mccall02b} conclude that `the chemistry that leads to H$_3^+$ is completely decoupled from that which is responsible for these heavier diatomics'. These factors lend support to the possibility that other molecules may be present in significant abundances in the more diffuse regions surrounding conventional molecular clouds.

Very rapid production mechanisms are required for molecules such as \cm\ and its anion to exist in the warm diffuse regions traced by \ion{Ti}{ii}, though PAHs and related large organic species will survive for much longer in the UV radiation field with the potential to give rise to diffuse interstellar bands. The interstellar media containing \ion{Ti}{ii} and \ion{Ca}{ii} are largely co-spatial \citep[see][]{crink94,albert93}, and these refractory ions are observed predominantly in the warm neutral ISM. \citet{welty96} found the mean upper temperature limit of the clouds in their \ion{Ca}{ii} survey to be 4100~K, though temperatures as low as a few hundred K were found in many cases. Titanium and calcium arise in the gas phase predominantly due to grain destruction \citep{stokes78}, and it is plausible that molecules and DIB carriers may also arise in this way.

There is no evidence in the literature for correlations between DIB strengths and \ion{Ti}{ii} column densities, though \citet{herbig93} found that the 5780 and 5797 DIBs do not correlate with the level of titanium depletion. This study suggests that further investigation of a possible link between DIBs, molecules and \ion{Ti}{ii} is warranted.

\section{Conclusion}

Using the rotational constants of \citet{lykke87}, the absorption spectrum arising from the \dbsgs\ transition of \cma\ has been computed for \cma\ in equilibrium with the CMB at 2.74~K.  The peak of the intrinsic absorption profile (at 8037.78~\AA) is consistent with the peak of the diffuse interstellar absorption feature at $8037.8\pm0.15$~\AA\ found in the spectra of eight heavily-reddened stars. If \cma\ occurs in the ISM with the statistical ortho~:~para abundance ratio of $3:1$ the \dbsgs\ transitions with $K_a''=1$ would produce strong spectral features at wavelengths of approximately $8024.8$ and $8049.6$~\AA\ that are not present in the observed interstellar spectra.  However, if the $K_a''=1$ level is populated based on a Boltzmann distribution at 2.74~K, then the $K_a=0\longleftarrow1$ and $K_a=2\longleftarrow1$ sub-bands are sufficiently weak to be compatible with their absence from the observed spectra. A thermal (Boltzmann) distribution of $K_a''$ levels would be expected if an efficient mechanism exists for conversion between the ortho and para forms of \cma.  Alternatively, the formation mechanism(s) for \cma\ may introduce a skewed (\ie\ non $3:1$) ratio of ortho and para forms.  Higher signal-to-noise ($\gtrsim2000$) spectroscopic observations of the \dib8037 region will assist in the search for the $K_a''=1$ sub-bands. Without detailed laboratory or theoretical studies, it is unclear whether or not the \cma\ formation mechanism(s) would result in the required deviation of the ortho-to-para ratio from the statistical value of $3:1$.

The cyanomethyl anion \cma\ is a plausible candidate for the carrier of the narrow diffuse interstellar band located at $8037.8$~\AA, provided a spectral line-broadening mechanism can be accounted for that produces an approximately Gaussian broadening corresponding to a Doppler $b$ of between 16 and 33~\kms\ which corresponds with the observed \ion{Ti}{ii} and \ion{Ca}{ii} profiles towards HD~183143 and HD~168112, and is comparable to the H$_3^+$ linewidths towards Cyg~OB2~12, Cyg~OB2~5 and HD~183143.

Measurement of the oscillator strength of the \dbsgs\ transition would enable the calculation of the interstellar column density of \cma\ that would be required to produce the \dib8037 DIB.  Assuming the transition is strong ($f\sim0.5$), and the \cm\ anion-to-neutral ratio is large ($\sim1$, as suggested by preliminary calculations of the anion formation and destruction rates), then towards Cyg OB2 12 the required average fractional abundance of neutral \cm\ relative to H$_2$ is $2\times10^{-11}$, which falls at the lower end of the observed range in dense clouds.

The spectroscopic consistency between the \dib8037 DIB and the calculated transitions of \cma\ cannot be considered an assignment, but provides strong motivation for further study of this molecular anion including a search for the pure rotational transitions of interstellar \cma.

\begin{acknowledgements}
The authors would like to thank George Herbig for kindly making available extensive Keck HIRES data and Ian Howarth for providing the \textsc{vapid} software and support. MAC thanks EPSRC for a studentship and The University of Nottingham for financial support.  We also thank Tom Millar and Steve Fossey for helpful comments, and a referee for suggested improvements to the manuscript. This research has made use of the SIMBAD database, operated at CDS, Strasbourg, France. The W. M. Keck Observatory is operated as a scientific partnership among the California Institute of Technology, the University of California, and NASA. The observatory was made possible by the generous financial support of the W. M. Keck Foundation.
\end{acknowledgements}

\bibliographystyle{7358aa}
\bibliography{7358refs}

\end{document}